\documentclass[aps,prl,twocolumn,showpacs,superscriptaddress,groupedaddress]{revtex4}
\usepackage{amsfonts}
\usepackage{amsmath}
\usepackage{amssymb}
\usepackage{graphicx}%
\providecommand{\U}[1]{\protect\rule{.1in}{.1in}}

\begin{document}
\preprint{ }
\title{Informatic error-disturbance relation in the qubit case}
\author{Li-Yi Hsu}
\affiliation{Department of Physics, Chung Yuan Christain University, Chungli, 320, Taiwan,
Republic of China}
\pacs{PACS number}

\begin{abstract}
In 1927, Heisenberg heuristically disclosed the tradeoff between the error in
the measurement and the caused disturbance on another complementary
observable. In quantum theory, most uncertainty relations are proposed to
describe the level of unavoidable uncertainty in the measurement process. In
this paper, we study the error-disturbance relation from an information
perspective. We ask how much information, rather than how much uncertainty,
can be gained during two sequential measurements. To achieve the optimal
information gain, we argue that the strategy for an "intelligent" prior
apparatus is to clone the unknown state and, for the posterior apparatus, the
swapping operation should be performed in the posterior apparatus. We propose
a coarse-grain random access code, and therein information causality as a
physical principle can be exploited to derive the upper-bound of information
gain. Finally, we conjecture the information gain of the position and momentum
using coarse-grain measurements.

\end{abstract}
\volumeyear{year}
\volumenumber{number}
\issuenumber{number}
\eid{identifier}
\date[Date text]{date}
\received[Received text]{date}

\revised[Revised text]{date}

\accepted[Accepted text]{date}

\published[Published text]{date}

\startpage{1}
\endpage{ }
\maketitle

As a fundamental principle in quantum physics, there is always tradeoff in
measuring two non-commuting observables. To access some observable of an
quantum object, one must specify definite experiments as the measurement
processes. In Heisenberg's original thought, determining an electron's
position with the attainable accuracy must reduce the measurement precision of
another complementary observable, such as momentum. In detail, when an
electron is scattered by photons, the position is instantly determined.
However, the momentum undergoes a "discontinuous change" due to the scattering
\cite{Note,2,3}. General speaking, in the sequential quantum measurements, the
prior measurement may include error, and the caused disturbance generates a
discontinuous change for the latter measurement outcome. Hence the term
"error-disturbance relation" is designated as a suitable name of uncertainty principle.

Researchers attempt to tackle quantum measurement uncertainty using
alternative approaches. Considering the thought experiment of observing an
electron using an imaginary gamma-ray microscope, Heisenberg put down the
relation $\bigtriangleup x\bigtriangleup p\sim h$ in 1927. Robertson's
well-known\ formula has been introduced in many standard quantum physics
textbooks, which states that the product of two standard deviations for two
non-commuting observables is upper-bounded \cite{Robertson}, therein the
error-and-disturbance scenario is poorly represented. On the other hand,
entropic uncertainty has been also investigated in \cite{en1,en2,en3,en4}. As
proposed by Busch, Lahti, and Werner (BLW), the Wasserstein distance between
probability distributions is exploited \cite{BLW2} and, potential experimental
schemes have been also proposed \cite{BLW1, BLW2, BLW3}. Recently, based on
the error-and-disturbance scenario, Ozawa's reformulation has attracted much
more attention \cite{o1,o2,o3}; therein, the error and disturbance is recalled
and operationally defined \cite{Oza qubit}. In particular, the qubit case with
discrete measurement outcomes \cite{Oza qubit,4} has been claimed to be
experimentally verified \cite{exp1,exp2}.

However, the Ozawa and BLW's approaches have been debated in the field
\cite{F1,F2,F3}, partially due to lack of\ well specification on the usage
\textquotedblleft uncertainty\textquotedblright. For Robertson's relation,
uncertainty implicitly indicates the product of two standard deviations;
however, the indication is not useful in the recent studies. On the other
hand, measurements are exploited to know the information of an object. A more
meaningful question is to ask how much information, rather than its
uncertainty, can be obtained in sequential measurements. Thanks to the quantum
information science, physicists can quantify information using mutual
information. Inevitable uncertainty never be removed, which implicitly
indicates the impossibility of accessing full information. In this paper, we
propose the upper bound for information that one can obtain from the
measurements of the two non-commuting observables.

Throughout we employ the von Neumann quantum measurement model, which can be
briefly stated as follows. The goal is to learn the information of an unknown
state of the quantum object $\mathbf{O}$ through sequential measurements. In
detail, a \textit{prior} measurement on the observable $A$ is performed using
apparatus $\mathbf{A}$, and a later measurement on the observable $B$ is
performed using apparatus $\mathbf{B}$. Before the readouts, the quantum
system $\mathbf{P}_{A}$ and $\mathbf{P}_{B}$ each as the probe or pointer in
the apparatus $\mathbf{A}$ and $\mathbf{B}$ interacts with the object
$\mathbf{O}$, respectively. The process is depicted in Figure 1. Hereafter the
object $\mathbf{O}$\ can be regarded as a qubit with binary measurement
outcomes. In addition, let the observables $A$ and $M_{A}$ ($B$ and $M_{B}$)
be the spin observables in the same direction.

Without loss of generality, the initial state of the composite system
$\mathbf{O}+\mathbf{P}$ is $\left\vert \psi\right\rangle \otimes\left\vert
0\right\rangle $ before turning on the interaction at time $t$. Notably,
$\left\vert \psi\right\rangle \ $is an unknown two-level state,
and\ $\left\vert 0\right\rangle $ is a fixed probe state; and these two states
are completely uncorrelated. At time $t+\bigtriangleup t$, the interaction is
turned off. $U$ denotes the unitary time evolution of $\mathbf{O}+\mathbf{P}$
during the interval ($t$, $t+\bigtriangleup t$). After the turn-off of the
interaction the sharp measurement on the observable $M\mathit{\ }$is
performed. In Ozawa's formulation \cite{OO1,OO2}, the noise value
$\epsilon(\psi)$ is defined as%
\[
\epsilon(\psi)=\left\langle \psi\otimes\xi\left\vert (M_{A}^{out}-A^{in}%
)^{2}\right\vert \psi\otimes\xi\right\rangle ^{1/2},
\]
where $M_{A}^{out}=U^{\dag}(I\otimes M_{A})U$ and $A^{in}=A\otimes I$. The
disturbance value is $\eta(\psi)$ defined as%

\[
\eta(\psi)=\left\langle \psi\otimes\xi\left\vert (B^{out}-B^{in}%
)^{2}\right\vert \psi\otimes\xi\right\rangle ^{1/2},
\]
where $B^{out}=U^{\dag}(B\otimes I)U$ and $B^{in}=B\otimes I$. For more
discussions on the limitation of operator $U$, readers can refer to
\cite{1408.2272}. For simplicity, we weaken the condition such that all kinds
of unitarity are feasible for the object-probe interactions.

\textit{ Case (a)} Let $U$ be the identity operator. Obviously, $B^{out}%
=B^{in}$; hence $\eta=0$. The object state $\psi$ is undisturbed, and no
information can be gained.

\textit{Case (b) }Let $U$ be the SWAP operation,%

\begin{equation}
U_{SWAP}\left\vert \psi\right\rangle \left\vert 0\right\rangle =\left\vert
0\right\rangle \left\vert \psi\right\rangle .
\end{equation}
As a result, we have%

\begin{equation}
\left\langle M^{out}\right\rangle =\left\langle A^{in}\right\rangle ,
\label{Exp}%
\end{equation}
and the error $\epsilon(\psi)=0$. The state of the object after the unitary
evolution is always fixed, which is completely uncorrelated with the initial
state. In this case, the state is regarded as mostly disturbed. As a result,
it is possible that $\epsilon(\psi)\eta(\psi)=0$.

The following remarks refer to Ozawa's scenario. First, Heisenberg's picture
is usually exploited in Ozawa's formulation. Here we exploit Schr\"{o}dinger's
picture, which aids in elucidating the substantial role of quantum cloning. We
can describe the sequential measurements as the quantum circuit, as shown in
Fig. 1(b). Second, as previously mentioned, we focus on information gain. How
the state is disturbed is not our main concern. We will show that error comes
from imperfect quantum cloning. Third, the concept of joint measurement is
exploited in Ozawa's study \cite{Oza qubit}. Nevertheless, the\ role of
respective unitary evolution in either apparatus $\mathbf{A}$ or $\mathbf{B}$
is not clear described.

The following interesting question naturally arises. Can $\epsilon=0$ and
$\eta=0$ simultaneously? The intuitive answer is positive with a $1\mapsto2$
cloning machine (PCM), that is,
\begin{equation}
U_{PCM}\left\vert \psi\right\rangle \left\vert 0\right\rangle =\left\vert
\psi\right\rangle \left\vert \psi\right\rangle . \label{U}%
\end{equation}
If one were to perfect $1\mapsto2$ cloning ($U_{A}=U_{PCM}$) followed by the
swapping operation between the object and probe $B$ ($U_{B}=U_{SWAP}$), the
states of both probes $P_{A}$ and $P_{B}$ would be the same state, $\left\vert
\psi\right\rangle $; hence it were to access full information of the measured
observables $A$ and $B$ both. Therefore, the quantum no-cloning prevents the
simultaneous vanishing of noise and disturbance. For physical realization, we
propose that the optimal information gain where $U_{A}$ as the optimal
$1\mapsto2$ cloner and $U_{B}$ as the swapping operation are used,
respectively. Notably, we do not know of other non-trivial or accidental means
for getting optimal information gain, which is outside the scope of this paper.

To tackle the error-disturbance relation from an informational perspective, we
propose the coarse-grained random access code (CRAC) with the following
scenario. Initially, Alice has a two-bit database $x_{A}x_{B}\in
\{00,01,10,11\}$, which the distant Bob would like to access. Alice and Bob
share an ensemble of the Bell state $\left\vert \Phi\right\rangle =\frac
{1}{\sqrt{2}}(\left\vert 0\right\rangle \left\vert 1\right\rangle -\left\vert
1\right\rangle \left\vert 0\right\rangle )$ as their physical resource. To
help Bob, Alice is permitted to perform classical one-bit communication with
Bob. Unlike the random access code in \cite{IC}, where Bob wants to access
only $x_{A}$ or $x_{B}$, he attempts to access both $x_{A}$ and $x_{B}$
simultaneously. In this case, Bob sequentially measures the observables
$A=\widehat{a}\cdot\overrightarrow{\sigma}$ and $B=\widehat{b}\cdot
\overrightarrow{\sigma}$, where\ $\overrightarrow{\sigma}=$ $(\sigma_{x}$,
$\sigma_{y}$, $\sigma_{z})$ and $\sigma_{x}$, $\sigma_{y}$, and $\sigma_{z}$
are Pauli matrix. Without loss of generality, the unit vectors $\widehat{a}$
and $\widehat{b}$ lie in the ($x$-$y$) equator of the Bloch sphere and divided
the equator into four quadrants $Q_{x_{A}x_{B}}$, as depicted in Figure 2.
Alice random chooses a unit Bloch vector $\widehat{\varphi}_{x_{A}x_{B}}\in
Q_{x_{A}x_{B}}$ with the angle $\varphi_{x_{A}x_{B}}$ in the equator. The
corresponding state of the phase $\varphi_{x_{A}x_{B}}$ is%
\[
\left\vert \varphi_{x_{A}x_{B}}\right\rangle =\frac{1}{\sqrt{2}}(\left\vert
0\right\rangle +e^{i\varphi_{x_{A}x_{B}}}\left\vert 1\right\rangle ),
\]
where the orthogonal state in the equator is $\left\vert \varphi_{x_{A}x_{B}%
}^{\perp}\right\rangle =\frac{1}{\sqrt{2}}(\left\vert 0\right\rangle
-e^{i\varphi_{x_{A}x_{B}}}\left\vert 1\right\rangle )$ \cite{Cop}. Up to a
global phase, the Bell state can be revised as
\[
\left\vert \Phi\right\rangle =\frac{1}{\sqrt{2}}(\left\vert \varphi
_{x_{A}x_{B}}\right\rangle \left\vert \varphi_{x_{A}x_{B}}^{\perp
}\right\rangle -\left\vert \varphi_{x_{A}x_{B}}^{\perp}\right\rangle
\left\vert \varphi_{x_{A}x_{B}}\right\rangle ).
\]

The bits $x_{w}=(1-\Theta(\widehat{\varphi}_{x_{A}x_{B}}\cdot\widehat{w}%
))\in\{0,$ $1\}$, where $w\in\{a,$ $b\}$ and the Heaviside step function%

\begin{align*}
\Theta(z)  &  =1,\text{ }z>0\\
&  =0,\text{ }z\leq0.
\end{align*}
The word \textquotedblleft coarse-grained\textquotedblright\ indicates Alice
can choose \textit{any} $\left\vert \varphi_{x_{A}x_{B}}\right\rangle $ to
encode her two-bit database and, she only concerns which quadrant includes
$\varphi_{x_{A}x_{B}}$. We state the proposed protocol as follows.

(1). Encoding phase: Alice perform the projective measurement with the
random-chosen orthonormal basis states \{$\left\vert \varphi_{x_{A}x_{B}%
}\right\rangle $, $\left\vert \varphi_{x_{A}x_{B}}^{\perp}\right\rangle $\}.

(2). Communication phase:\textit{ }Alice announces the classical bit $\beta=1$
($0$) if the post-selected state is $\left\vert \varphi_{x_{A}x_{B}%
}\right\rangle $ ($\left\vert \varphi_{x_{A}x_{B}}^{\perp}\right\rangle $).

(3). Decoding phase:\textit{ }As shown in Fig. 1, Bob sequentially performs
the sharp measurement $A$\ and $B$ with the outcomes $O_{A}$and $O_{B}%
\in\{1,-1\}$, respectively. Finally Bob's guessing answers on $x_{A}$ and
$x_{B}$ are $g_{A}=O_{A}^{\prime}+\beta$ (mod2) and $g_{B}=O_{B}^{\prime}$
$+\beta$ (mod2), where $O_{A}^{\prime}=\frac{1-O_{A}}{2}$ and $O_{B}^{\prime
}=\frac{1-O_{B}}{2}\in\{0,1\}$, respectively.

According to information causality \cite{IC, IC1}, we have
\begin{equation}%
{\displaystyle\sum\nolimits_{W\in\{A,B\}}}
I(x_{W}:g_{W})\leq1, \label{IC}%
\end{equation}
where $I(x:g)$ denotes the mutual information between the random variables $x$
and $g$. Let the state $\left\vert \varphi_{x_{A}^{\prime}x_{B}^{\prime}%
}\right\rangle $ be Bob's qubit state after Alice's local measurement.
Ineq.(\ref{IC}) can be revised as follows%

\begin{equation}%
{\displaystyle\sum\nolimits_{W\in\{A,B\}}}
I(x_{W}:o_{W}^{\prime})\leq1. \label{main}%
\end{equation}

To evaluate the upper bound of the mutual information, we exploit the
following lemma.

\textit{Evan-Schulman} \textit{Lemma :} Consider a cascade of two
communication channels: X $\hookrightarrow$Y $\hookrightarrow$ Z, with $X$,
$Y$, $Z$ being random variables. Let $Y$ and $Z$ be the input and output of
the symmetric channel $C$, respectively, with the successful transmission
probability $\frac{1+\xi}{2}$. \ We have
\begin{equation}
\frac{I(X;Z)}{I(X;Y)}\leq\xi^{2}. \label{ratio}%
\end{equation}
Interested readers can refer to \cite{es,es2} for the detailed rigorous proof.

Without loss of generality, assume the bias parameter $0\leq\xi\leq1$, and
hereafter let the $X$ and $Y$ be the input and output of an error-free channel
and hence we have $I(X;Y)=1$. Let variables $Y$ and $Z$ be $x_{w}^{\prime}$
and $o_{w}^{\prime}$, respectively. Finally we have
\begin{equation}
I(x_{w};\text{ }o_{w}^{\prime})\leq\xi_{w}^{2}. \label{II}%
\end{equation}

To achieve the optimal encoding, the unitary operator $U_{A}$ corresponds to
the phase covariant $1\mapsto2$ cloner (PCC), which reads \cite{clone, C1,
C2,C3}
\begin{equation}
U_{PCC}\left\vert 0\right\rangle \left\vert 0\right\rangle =\left\vert
0\right\rangle \left\vert 0\right\rangle , \label{c1}%
\end{equation}
and%

\begin{equation}
U_{PCC}\left\vert 1\right\rangle \left\vert 0\right\rangle =\cos\eta\left\vert
1\right\rangle \left\vert 0\right\rangle +\sin\eta\left\vert 0\right\rangle
\left\vert 1\right\rangle . \label{c2}%
\end{equation}
Specifically, no ancilla qubit is required in the cloning process.

Now we can state the main result for this paper as follows. According to
(\ref{main}) and (\ref{II}), the information gain
\begin{equation}
I=%
{\displaystyle\sum\nolimits_{W\in\{A,B\}}}
I(x_{W}^{\prime}:o_{W}^{\prime})\leq\xi_{A}^{2}+\xi_{B}^{2}\leq1,
\label{Result}%
\end{equation}
where
\begin{equation}
\xi_{A}=\left\vert \widehat{a}\cdot\widehat{\varphi}_{x_{A}x_{B}}\right\vert
\sin\eta,\text{ and }\xi_{B}=\left\vert \widehat{b}\cdot\widehat{\varphi
}_{x_{A}x_{B}}\right\vert \cos\eta. \label{bias}%
\end{equation}
The details for calculating $\xi_{A}$ and $\xi_{B}$\ are given in the
Supplemental material. Notably, the bias parameters each are the products of
two parts. One part is from the cloning coefficients ($\sin\eta$, $\cos\eta$),
and the other is from the measurement process ($\left\vert \widehat{a}%
\cdot\widehat{\varphi}_{x_{A}x_{B}}\right\vert ,\left\vert \widehat{b}%
\cdot\widehat{\varphi}_{x_{A}x_{B}}\right\vert $ ). We consider the following
specific cases.

Case (a) Consider either $\xi_{A}=1$ or $\xi_{B}=1$. To fulfill $\xi_{A}=1$
($\xi_{B}=1$), the equation $\left\vert \sin\eta\right\vert =\left\vert
\widehat{a}\cdot\widehat{\varphi}_{x_{A}x_{B}}\right\vert =1$ ($\left\vert
\cos\eta\right\vert =\left\vert \widehat{b}\cdot\widehat{\varphi}_{x_{A}x_{B}%
}\right\vert =1)$ must hold; hence $\xi_{B}=\cos\eta=0$ ($\xi_{A}=\sin
\eta=0).$ Here, $U_{A}$ is only a swapping operator (an identity operator).
Furthermore, $\widehat{a}$ ($\widehat{b}$) and the unknown $\widehat{\varphi
}_{x_{A}x_{B}}$ are accidentally in either the same or opposite direction. In
CRAC, when $\left\vert \widehat{a}\cdot\widehat{\varphi}_{x_{A}x_{B}%
}\right\vert =1$ ($\left\vert \widehat{b}\cdot\widehat{\varphi}_{x_{A}x_{B}%
}\right\vert =1$)$,$ Alice and Bob's outcomes must either perfect or
anti-perfect correlated, which leads to 1-bit of information gain.

Case (b) Consider the symmetric cloning with $\eta=\frac{\pi}{4}$. The
condition $\left\vert \widehat{a}\cdot\widehat{\varphi}_{x_{A}x_{B}%
}\right\vert =\left\vert \widehat{b}\cdot\widehat{\varphi}_{x_{A}x_{B}%
}\right\vert =1$ yields 1-bit of information gain. However, the latter
measurement is meaningless because $\widehat{b}$ is either parallel or
anti-parallel to $\widehat{a}$.

Case (c) Let the observables of these two measurements be most incompatible
(e.g., $\widehat{a}\cdot\widehat{b}=0$). In this case, we set $\widehat
{a}\cdot\widehat{\varphi}_{x_{A}x_{B}}=\cos\delta$ and $\widehat{b}%
\cdot\widehat{\varphi}_{x_{A}x_{B}}=\sin\delta$. The optimal value $I$ in
(\ref{Result}) can be achieved using the symmetric cloning with the condition
$\delta=\frac{\pi}{4}$. As a result,
\begin{equation}
I\leq\frac{1}{2}. \label{MM1}%
\end{equation}

In the general CRAC, Alice has an $N$-bit local database $\overrightarrow
{x}=x_{1}\ldots x_{2}$, where $x_{i}=(1-\Theta(\widehat{\varphi}%
_{\overrightarrow{x}}\cdot\widehat{m}_{i}))$ $\forall$ $i$, and the
measurement basis states \{$\left\vert \varphi_{\overrightarrow{x}%
}\right\rangle $, $\left\vert \varphi_{\overrightarrow{x}}^{\perp
}\right\rangle $\} are exploited during the encoding phase followed by
classical one-bit communication. Similarly, Bob wants to access the $i$-th bit
$x_{i}$ by measuring $\widehat{m}_{i}\cdot\overrightarrow{\sigma}$ with
$\forall i=1,\ldots,N$. The proposed optimal strategy for Bob is to performs
$1\mapsto N$ phase covariant cloning and measures the observable $\widehat
{m}_{i}\cdot\overrightarrow{\sigma}$ on the object with the $\mathit{i}$-th
copied state. Thus the information gain from measuring $\widehat{m}_{i}%
\cdot\overrightarrow{\sigma}$ is $I_{i}\leq$ $\xi_{m_{i}}^{2}$. Based on
(\ref{bias}), we conjecture that the $i$-th bias parameter
\[
\xi_{m_{i}}=(\widehat{e}_{i}\cdot\overrightarrow{n})(\widehat{m}_{i}%
\cdot\widehat{\varphi}_{\overrightarrow{x}}),
\]
where $\{\widehat{e}_{1},\ldots\widehat{e}_{N}\}$ is a set of the orthonormal
basis vectors in the $N$ dimensional space $\mathit{R}^{N}.$ The vector
$\overrightarrow{n}\in R^{N}$ and $\left\vert \overrightarrow{n}\right\vert
\leq1$. Here the vectors $\widehat{e}_{1},\ldots,\widehat{e}_{N}$ and
$\overrightarrow{n}$ \ should be parameterized according to the $1\mapsto N$
cloning process. Such conjecture ensures that preservation of information
causality,
\[%
{\displaystyle\sum\limits_{i=1}^{N}}
I_{i}\leq1.
\]
Notably, only $1\mapsto2\ $phase covariant cloning is \textquotedblleft simple
enough\textquotedblright\ that can be done using the single pair-wise
interaction between the object and the probe. As for $1\mapsto N$ cloning
process with $N>2$ should involve more quantum objects. In this case, the
quantum cloning as the pre-process should be performed before the any measured
object enters into any apparatus.

\begin{figure}[t]
\includegraphics[width=8cm]{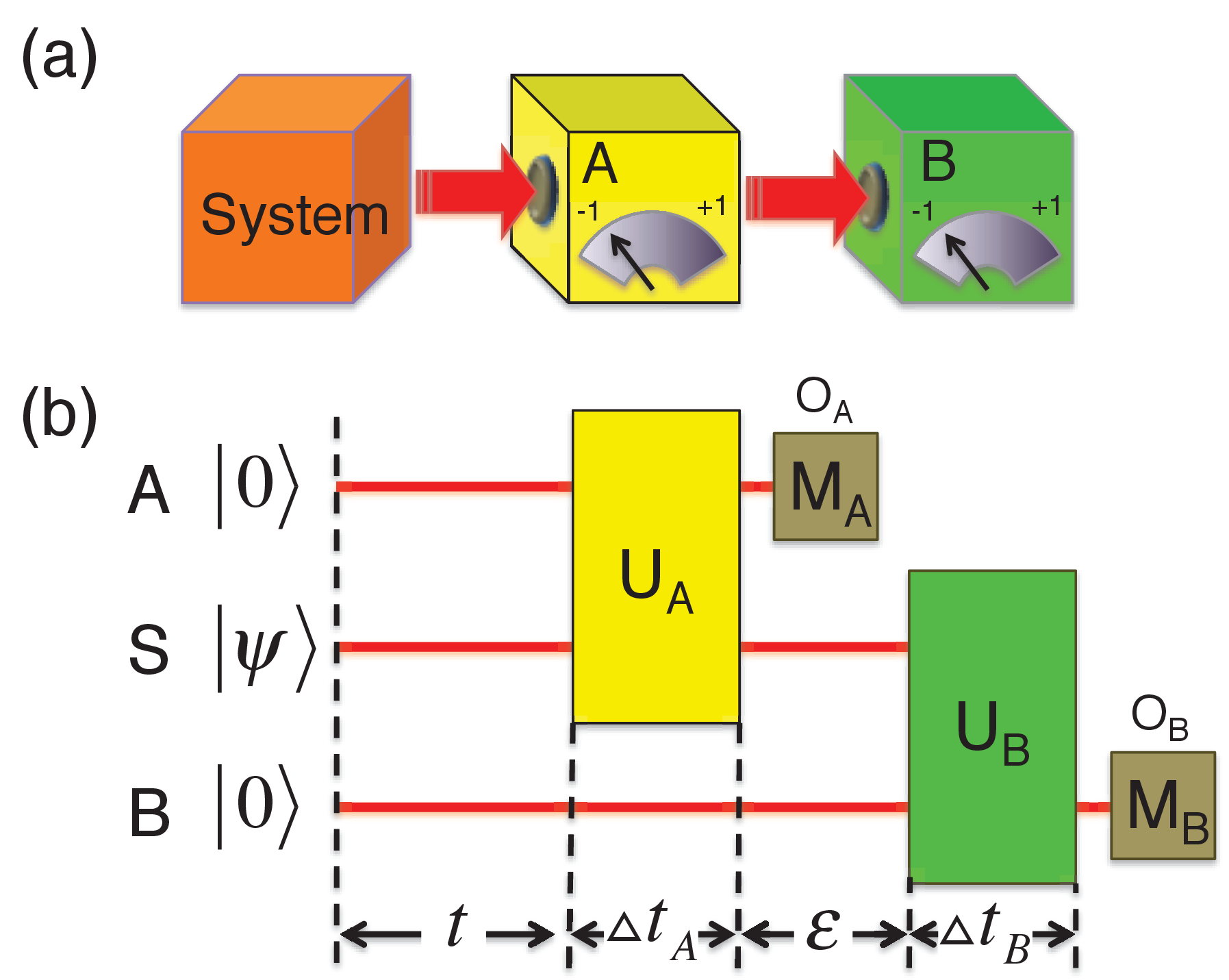}
\caption{(color online). (a) (b) The whole unitary operator $U$ \ can be decomposed as
$U=U_{B}U_{A}$. Firstly, during the time interval [$t,t+\bigtriangleup t_{A}$]
([$t_{B}=t+\bigtriangleup t_{A}+\varepsilon,$ $t_{B}+\bigtriangleup t_{B}$]),
the evolutions of the pair-wise interactions between the object and the probe
$\mathbf{P}_{A}$ ($\mathbf{P}_{B}$) is represented by unitary operator $U_{A}$
($U_{B}$). The initial states of the probes $\mathbf{P}_{A}$ and
\ $\mathbf{P}_{B}$ are $\left\vert 0\right\rangle _{A}$ and $\left\vert
0\right\rangle _{B}$, respectively. To gain optimal information, $U_{A}$
should be the optimal $1\mapsto2$ cloner, and $U_{B}$ the swapping operator.
}\label{image}
\end{figure}

\begin{figure}[t]
\includegraphics[width=8cm]{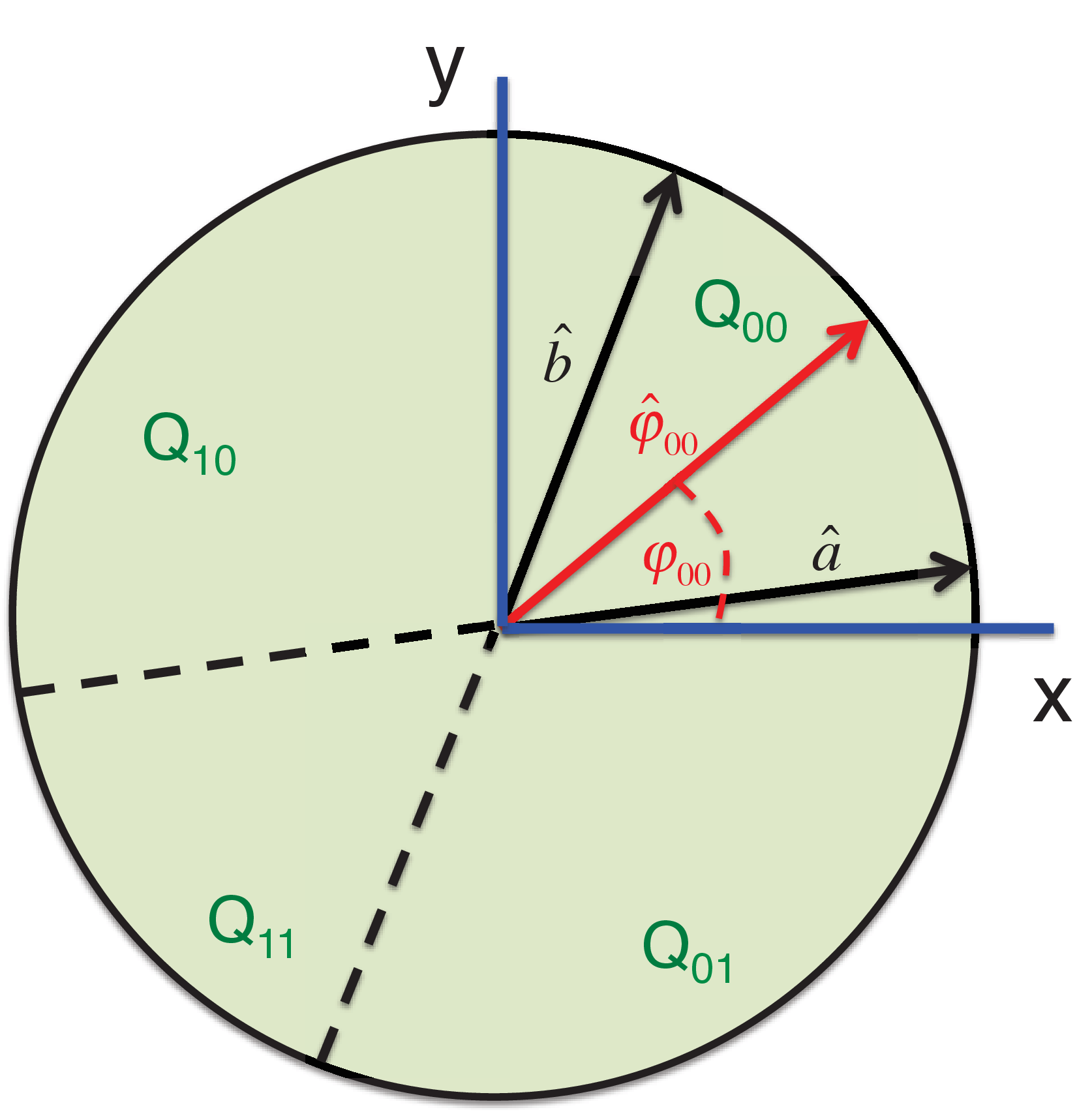}
\caption{(color online). The two dashed lines are the extensions of the unit vectors
$\widehat{a}$ and $\widehat{b}$, which divide the Bloch sphere ($x$-$y$)
equator plane four unequal quadrants: $Q_{00}$, $Q_{01}$, $Q_{10}$, and
$Q_{11}$.
}\label{image}
\end{figure}

At the end of the paper, we proposed a method of employing the CRAC in the
continuous variables, such as position and momentum of a one-dimensional
quantum system. Instead of precisely measuring the position and momentum, we
suppose there are two coarse-grained measurements. Where one is performed to
answer \textquotedblleft Is the\ object \textit{at left }or \textit{at
right}?\textquotedblright, and the other is performed to
answer\ \textquotedblleft Does the\ object \textit{go left }or \textit{goes
right}?\textquotedblright\ As a famous example, the physical realization of
answering the prior question can be regarded as the \textquotedblleft which
way\textquotedblright\ measurement in the double-slit experiment
\cite{WhichWay}. On this issue, Bohr argued that a measurement capable of
definitively discerning two positions (one-bit information) must produce an
\textquotedblleft uncontrollable change in the momentum\textquotedblright%
\ (zero information) \cite{Bohr}. Therefore, at most one-bit information can
be gained. On the other hand, the corresponding observables for the
coarse-grained position and momentum should be less incompatible. According to
(\ref{MM1}), the information gain can be greater than 0.5 bit. Let $I_{pm}$ be
the information gain from the coarse-grained position-and-momentum experiment.
Based on the above argument, we conjecture that
\[
\frac{1}{2}\leq I_{pm}\leq1.
\]

In summary, we propose an alternative way of studying sequential quantum
measurement from the infomatic perspective. The imperfect quantum cloning and
information causality are key to determining the upper bound of the
information gain. We also show that the fidelity of quantum cloning is limited
by information causality. Anyway, we only propose a toy model. The constraint
on time evolution unitary between the object and either probe should be
seriously considered for further study.

We acknowledge the financial support from Ministry of Science and Technology
of the Republic of China under Contracts NSC.102-2112-M-033-006-MY3.

\section{Supplemental material}

Without loss of generality, the spin observables along the directions of the
unit vectors $\widehat{a}=(a_{x},$ $a_{y}$, $0)$ and $\widehat{b}=(a_{x},$
$a_{y}$, $0)$ are sequentially measured using the apparatus $\mathbf{A}$ and
$\mathbf{B}$, respectively. The sharp projectors $\Pi_{\pm}^{A}=\frac{1}%
{2}(I\pm\widehat{a}\cdot\overrightarrow{\sigma})$ ($\Pi_{\pm}^{B}=\frac{1}%
{2}(I\pm\widehat{b}\cdot\overrightarrow{\sigma})$) are exploited to measure
the observable $A$ ($B$). Let the probe state of apparatus $\mathbf{A}$ be
$\left\vert 0\right\rangle _{A}$. After the unitary operation $U_{A}$ as the
phase covariant $1\mapsto2$ cloning, the density matrix of the probe $A$ is
\[
\rho_{A}=tr_{S}\{\left\vert \psi\right\rangle \left\langle \psi\right\vert
\},
\]
where
\begin{align*}
\left\vert \psi\right\rangle  &  =U_{A}\left\vert \varphi_{x_{A}x_{B}%
}\right\rangle _{S}\left\vert 0\right\rangle _{A}\\
&  =\frac{1}{\sqrt{2}}(\left\vert 0\right\rangle \left\vert 0\right\rangle
+e^{i\varphi_{x_{A}x_{B}}}(\cos\eta\left\vert 1\right\rangle \left\vert
0\right\rangle +\sin\eta\left\vert 0\right\rangle \left\vert 1\right\rangle
)).
\end{align*}
The measurement outcome $o_{A}\in\{-1,1\}$ can be obtained with the
probability
\begin{align}
p_{A}  &  =tr(\rho_{A}\Pi_{sgn(o_{A})}^{A})\nonumber\\
&  =\frac{1}{2}(1+o_{A}(\widehat{a}\cdot\widehat{\varphi}_{x_{A}x_{B}}%
)\cos\eta). \tag{S1}%
\end{align}
Virtually, if $sgn(o_{A})=sgn(\widehat{\varphi}_{x_{A}x_{B}}\cdot\widehat{a}%
)$, the decoding is successful under the mapping on $o_{A}$ : $1\rightarrow0$
and $-1\rightarrow1$. According to (S1), the successful decoding probability
$p_{A}=\frac{1}{2}(1+\left\vert \widehat{a}\cdot\widehat{\varphi}_{x_{A}x_{B}%
}\right\vert \cos\eta)=\frac{1}{2}(1+\xi_{A})$. Therefore,%
\[
\xi_{A}=\left\vert \widehat{a}\cdot\widehat{\varphi}_{x_{A}x_{B}}\right\vert
\cos\eta.
\]

Next, after the unitary evolution $U_{B}$, the state of the probe $P_{B}$ is
swapped with that of the object. Hence the density matrix of the probe is
\[
\rho_{B}=tr_{A}\{\left\vert \psi\right\rangle \left\langle \psi\right\vert
\}.
\]
Using the above similar calculation and argument, we have the biased parameter%
\[
\xi_{B}=\left\vert \widehat{b}\cdot\widehat{\varphi}_{x_{A}x_{B}}\right\vert
\sin\eta.
\]

\bigskip

\begin{thebibliography}{99}                                                                                               %


\bibitem {Note}Bohr gave a full account of measuring of an electron using a
$\gamma$-ray microscope, which is published in \cite{2}. Heisenberg finally
referred to Bohr's scenario instead of his own. Interesting readers can refer
to \cite{3}.

\bibitem {2}N. Bohr, Nature\textbf{ 128}, 580 (1928).

\bibitem {3}K. Camilleri, \textit{Heisenberg and the interpretation of quantum
mechanics}, (Cambridge University Press, Cambridge, 2011).

\bibitem {Robertson}H.P. Robertson, Phys. Rev. \textbf{34}, 163 (1929).

\bibitem {en1}Bia lynicki-Birula, I., Mycielski, J, Commun. Math. Phys.
\textbf{44}, 129 (1975).

\bibitem {en2}D. Deutsch, Phys. Rev. Lett. \textbf{50}, 631 (1983).

\bibitem {en3}H. Maassen, J. B. M. Uffink, Phys. Rev. Lett. \textbf{60}, 1103 (1988).

\bibitem {en4}F. Buscemi, M. J.\thinspace W. Hall, M. Ozawa, and M. M. Wilde,
Phys. Rev. Lett. \textbf{112}, 050401 (2014).

\bibitem {BLW2}P. Busch, P. Lahti, and R. F. Werner, Phys. Rev. A \textbf{89},
012129 (2014).

\bibitem {BLW1}P. Busch, P. Lahti, and R. F. Werner, Phys. Rev. Lett.\textbf{
111}, 160405 (2013).

\bibitem {BLW3}P. Busch, P. Lahti, and R. F. Werner, arXiv:1312.4393.

\bibitem {o1}M. Ozawa, Phys. Lett. A \textbf{320}, 367 (2004).

\bibitem {o2}M. Ozawa, Ann. Phys. (N.Y.) \textbf{311}, 350 (2004).

\bibitem {o3}M. Ozawa, J. Opt. B: Quantum Semiclass. Opt. \textbf{7}, S672 (2005).

\bibitem {Oza qubit}M. Ozawa, Phys. Rev. A \textbf{67}, 042105 (2003).

\bibitem {4}A. P. Lund and H. M. Wiseman, New J. Phys. \textbf{12}, 093011 (2010).

\bibitem {exp1}J. Erhart, S. Sponar, G. Sulyok, G. Badurek, M. Ozawa, and Y.
Hasegawa, Nature Phys. \textbf{8}, 185, (2012).

\bibitem {exp2}L.A. Rozema, A. Darabi, D.H. Mahler, A. Hayat, Y. Soudagar, and
A.M. Steinberg, Phys. Rev. Lett. \textbf{109}, 100404 (2012).

\bibitem {F1}M. Ozawa, arXiv:1308.3540.

\bibitem {F2}P. Busch, P. Lahti, and R. F. Werner, arXiv:1402.3102.

\bibitem {F3}P. Busch, P. Lahti, and R. F. Werner, arXiv:1403.0367.

\bibitem {OO1}M. Ozawa, Int. J. Quant, Inf. \textbf{1}, 569 (2003).

\bibitem {OO2}M. Ozawa, arXiv:1402.5601.

\bibitem {1408.2272}R. Silva, N. Gisin, Y. Guryanova, S. Popescu, arXiv:1408.2272.

\bibitem {IC}M. Pawlowski, T. Paterek, D. Kaszlikowski, V. Scarani, A. Winter,
M. Zukowski, Nature \textbf{461}, 1101 (2009).

\bibitem {Cop}As a result, three unit vectors $\widehat{a}$, $\widehat{b}$,
and $\widehat{\varphi}_{x_{A}x_{B}}$ are assumed coplanar. Any $\widehat
{\varphi}_{x_{A}x_{B}}$ lying out of the equator can only reduce the values
$\left\vert \widehat{a}\cdot\widehat{\varphi}_{x_{A}x_{B}}\right\vert $ and
$\left\vert \widehat{b}\cdot\widehat{\varphi}_{x_{A}x_{B}}\right\vert $ in
(\ref{Result}), and hence the information gain.

\bibitem {IC1}L.-Y. Hsu, I-C. Yu, and F.-L. Lin, Phys. Rev. A \textbf{84},
042319 (2011).

\bibitem {es}W. Evans and L. J. Schulman, in Proceedings of the 34th Annual
Symposium on Foundations of Computer Science (IEEE, New York, 1993), p. 594.

\bibitem {es2}W. Evans and L. J. Schulman, IEEE Trans. Inf. Theory
\textbf{45}, 2367 (1999).

\bibitem {clone}V. Scarani, S. Iblisdir, N. Gisin, and A. A\'{c}in, Rev. Mod.
Phys. \textbf{77}, 1225 (2005).

\bibitem {C1}C.-S. Niu and R. B. Griffiths, Phys. Rev. A \textbf{60}, 2764 (1999).

\bibitem {C2}T. Durt and J. Du, Phys. Rev. A \textbf{69}, 062316 (2004).

\bibitem {C3}D. Bru$\beta$ , M. Cinchetti, G. M. D'Ariano, and C.
Macchiavello, Phys. Rev. A \textbf{62}, 012302 (2000).

\bibitem {WhichWay}Feynman, R. P., Leighton, R. B. \& Sands, M. \textit{The
Feynman Lectures on Physics Vol. III} (Addison Wesley, Reading MA, 1965).

\bibitem {Bohr}N. Bohr, Naturwissenschaften, 16, 245 (1928).
\end{thebibliography}
\end{document}